\documentclass[conference]{IEEEtran}
\usepackage[noadjust]{cite}

\ifCLASSINFOpdf
  \usepackage[pdftex]{graphicx}
  \graphicspath{{../pdf/}{../jpeg/}}
\else
\fi
\usepackage[cmex10]{amsmath}
\usepackage[table,xcdraw]{xcolor}

\usepackage{algorithm,algorithmic}

\usepackage{mdwmath}
\usepackage{mdwtab}

\usepackage{booktabs}

\begin{document}
%
\title{S-CHIRP: Secure Communication for\\ Heterogeneous IoTs with Round-Robin Protection}

\author{\IEEEauthorblockN{Mike Borowczak, \textit{Member, IEEE}} 
\IEEEauthorblockA{Computer Science Department\\
University of Wyoming\\
Laramie, Wyoming 82071, USA\\
Email: Mike.Borowczak@uwyo.edu}
\and
\IEEEauthorblockN{George Purdy}
\IEEEauthorblockA{Computer Science Department\\
University of Cincinnati\\
Cincinnati, Ohio 45221, USA\\
Email: George.Purdy@uc.edu}}


%

\makeatletter
\def\imod#1{\allowbreak\mkern10mu({\operator@font mod}\,\,#1)}
\makeatother

\maketitle

\begin{abstract}
This work introduces CHIRP - an algorithm for communication between ultra-portable heterogeneous IoT devices with a type of round-robin protection mechanism. This algorithm is presented both in its basic form as well as in a secured form in order to secure and maintain trust boundaries and communication within specific groups of heterogeneous devices. The specific target application scenarios includes resource constrained environments where a co-located swarm of devices (adversarial in mission or objective) is also present. CHIRP, and its secured version (S-CHIRP), enables complete peer-to-peer communication of a $n$-agent network of devices in as few as $n$ rounds. In addition to the n-round cycle length, the proposed communication mechanism has the following major properties: nodes communication is entirely decentralized, communication is resilient to the loss of nodes, and finally communication is resilient to the (re)-entry of nodes. Theoretical models show that even the secure implementation of this mechanism is capable of scaling to IoT swarms in the million device range with memory constraints in the $< 10$ MB range.

\end{abstract}


%
\IEEEpeerreviewmaketitle

\section{Introduction}
As the Internet of Things (IoT) moves from a sparse ecosystem of point-to-endpoint connected devices to a densely distributed mesh network, one of the biggest challenges future designers face is how to maintain trust boundaries and communication within specific groups of heterogeneous devices in resource constrained environments while still allowing untrusted / unknown communication to pass through devices unabated. This challenge is complicated with the adoption of ultra-portable (low-power, low-compute) IoT devices. 

Consider the following scenario - a swarm of several hundred (thousand) micro-autonomous vehicles are released to surround a set of targets (e.g. tactical, nuclear accidents, e.t.c.), simultaneously another co-located swarm is released (adversarial in mission or objective). The following questions drive this work:
\begin{enumerate}
\item How can we ensure separation between agents of these swarms (even if one of the swarms is intentionally adversarial in nature)? 
\item Perhaps of greater significance, can this swarm communicate in resource constrained / hostile / contested environments such as when placed within a protected/blocked electromagnetic (EM) environment (i.e. saturation of $\geq$ 20 THz spectrum)?
\item Finally, can swarm agents by reprogrammed dynamically to a communicate with a new swarm (i.e. software defined association / security)?
\end{enumerate}

This work aims to provide a solution to this problem in the form of a lightweight protocol for \textbf{S}ecure-\textbf{C}ommunication for \textbf{I}oTs with \textbf{R}ound-Robin \textbf{P}rotection (S-CHIRP). S-CHIRP enables complete peer-to-peer communication of an $n$-agent network of devices in as fews as $n$ rounds. Implementation of such a communication protocol can leverage lower frequencies within EM spectrum ($<$ 20 THz) including the visible light spectrum. Additionally, should the entire spectrum be saturated - communication could occur through a physical medium in the form of mechanical wave (e.g. sound) without modification to the underlying S-CHIRP protocol. Furthermore, this foundation of this work allows for other security mechanisms (e.g. peer-to-peer encryption strategies) while still leveraging the fundamental communication protocol solution (CHIRP).

\subsection{Related Work}
A significant amount of work has already been proposed to leveraging and enhancing existing technology to quickly enable the predicted exponential growth of IoT devices. This work focuses a security-centric approach to several open-research areas within the IoT space, namely\textit{Mobility Support} and \textit{Authentication}.\cite{ATZORI20102787}. 

Within the first domain of Mobility Support, most of the present day work focuses on enabling existing addressing protocols (IPv4 or IPv6) within a variety of solutions including RFID \cite{6550454, 4624080, 7571826} 6LoWPAN \cite{7997471}, and ROLL \cite{7745304}. The biggest issues with protocol reuse is the cost required to create \textit{adaptable heterogeneous networks} - the addition, replacement, loss or removal of IoT nodes becomes increasingly complex and requires centralized monitoring and control, as well as per-agent trust. Node to node communication is generally not feasible without storage of complete network topology knowledge.

Work within the Authentication space of IoTs has been significant - mainly in mechanisms to create and distribute keys efficiently and securely throughout a network \cite{CASTIGLIONE2017313,MESSAI201660,ZHOU2015255}. Most of these solutions face significant issues with adaptable heterogenous networks where the nodes are ephemeral - either due to movement physical movement out of the network, addition to the environment, or destruction via some outside force. Also almost all solutions are also susceptible to the the Proxy Attack problem - known as Man In Middle attacks \cite{ATZORI20102787} due to the predictability of communication patterns within Peer-to-Peer networks and the innate nature of Asymmetric communications. This work enables resuse of key-sharing and creation strategies, but over a flexible, adaptable and dynamic peer-to-peer network. 

\subsection{Contributions}
This work contributes:
\begin{enumerate}
\item An unmanaged, all-to-all, direct peer-to-peer communication protocol which completes in as little as $n$ rounds for an $n-$node network (CHIRP),
\item A mechanism to enhance CHIRP with per-network global keys such that individual nodes can still detect rouge communication attempts (S-CHIRP),
\item An implementation of the protocol on synthetic IoT swarm agents using a multi-agent simulator.
\end{enumerate}

\subsection{Outline of the Paper}
The remainder of the paper is structured as follows: Section \ref{CHIRP} explains the communication model and the overarching mechanism that allows for the unmanaged bi-directional communication of $n-$nodes in as little as $n$ rounds. The section also includes security considerations; and finally, Section \ref{FUTURE} discusses broader implications and future work.

\section{CHIRP}\label{CHIRP}

In order to develop a communication protocol capable of working across heterogeneous, ultra-portable, IoT devices the following constraints are considered:  
\begin{enumerate}
\item Centralized control of devices is not feasible,
\item a fluid topology does not allow for optimization of broadcasting,
\item Memory / processing limitations prevent retention, and distribution of aggregated packets,
\item As the scale (number of nodes $n$ increases) per-node keys are not-feasible,
\item Devices may leave and rejoin the network (e.g. loss of device, new device replacement, future network shrinkage/growth expected).
\end{enumerate}

The CHIRP solution presented here addresses these items by considering a network of $n$ IoT agents, $A =\left\lbrace a_1, a_2, \cdots , a_n  \right\rbrace$, and a set of $m$ rouge/adversarial agents, $R = \left\lbrace r_1, r_2, \cdots , r_m  \right\rbrace$. The generic communication model only considers the set of known agents $A$, while the section on enabling secure communication uses both sets of agents $A$ and $R$.

\subsection{Generic Communication Model}
The goal of CHRIP is to create an unmanaged, all-to-all, direct peer-to-peer communication protocol which completes in as little as $n$ rounds for an $n-$node network. Based on the constraints listed earlier the following four requirements are enumerated for the basic communication mode:

\begin{enumerate}
\item A node must communicate with every other node in the system in $n$ rounds (1 cycle), where $n$ is the maximum number nodes expected in the system ($\text{Node}_{cnt}$);
\item A node must communicate with at most one other node in any given round using only self-contained information;
\item Communication in the system must be resilient to the loss of nodes;
\item Communication in the system must be resilient to the addition of new nodes, so long as the total number of nodes is less than $\text{Node}_{cnt}$ and the index of the node is within the range $\left[ 0, \text{Node}_{cnt}\right)$ and has no collisions with existing nodes.

\end{enumerate}

\subsubsection{Communication within N-Rounds}
Given a set $A$ of $n$ agents, map these $n$ agents to a graph $G$ consisting of $n$ distinct nodes. Let the set of nodes belonging to $G$ be numbered from 1 to $n$ such that $G = \left\lbrace n_1, n_2, \cdots n_n \right\rbrace$. The objective CHIRP can be reduced to the following problem: 
Given a fully disconnected graph ($G$) of $n$ nodes, create a fully connected graph of $\frac{n(n-1)}{2}$ edges ($e$) within $n$ rounds. In order to accomplish this with direct peer-to-peer communication during any one round an edge ($e_{xy} \equiv e_{yx}$) is formed when two nodes ($n_x$, $n_y$) are jointly paired and $x \neq y$. 

The maximum number of edges ($E_{max}$) that can be formed in a single round is $\lfloor n/2 \rfloor$. Explicitly, if $n$ is even, $E_{max} = \lfloor n/2 \rfloor =  n/2 $. If, however, $n$ is odd, then $E_{max} = \lfloor n/2 \rfloor =  (n-1)/2 $. Thus, as shown in equation \ref{eq:1}, two distinct cases exist for the theoretical lower bound for the number of rounds needed to generate a complete graph ($R_{min}$). 

\begin{equation}  \label{eq:1}
  R_{min} = \begin{cases}
  \dfrac{\frac{n(n-1)}{2}}{\frac{n}{2}} = n-1 &\text{$n$ even}\\
  \dfrac{\frac{n(n-1)}{2}}{\frac{n-1}{2}} = n &\text{$n$ odd}
  \end{cases}
\end{equation}

\subsubsection{Per-Round Bi-directional Communication}
Through the remainder of this work the minimum number of rounds needed to accomplish complete pair-wise communication reverts to the worst case: $R_{min} = n$. CHIRP completes pairwise communication within $n$ rounds, where each round contains at least $\frac{n-1}{2}$ distinct pairs of communicating nodes. In order to accomplish distinct pairing a mechanism to determine node pairs during any given round must be defined. In order to determine the index of a source node's "target node pair" ($target_{idx}$) the following must known by the source node:
\begin{LaTeXdescription}
\item[Node Count ($\textit{Node}_{cnt}$) -] The maximum possible number of nodes within the system: $\textit{Node}_{cnt} \geq n$,
\item[Source Index ($\textit{self}_{idx}$) -] The source node's own index: $idx = \left[0,n\right)$
\item[Current Round ($\textit{r}_i$) - ] The current round of communication: $r = \left[0,n\right)$
\end{LaTeXdescription}

With these three parameters - any node can be paired with any other node using simple modular arithmetic during any round using equation \ref{eq:pairs}. Intuitively, a node's target pair is computed by subtracting its index from the current round while insuring that the result remains within the set of allowable nodes ($\bmod \textit{ Node}_{cnt}$). Imagine the current round as a mechanism to ensure that a node can sweep through the nodes at fixed distances from itself.

\begin{equation}\label{eq:pairs}
\textit{self}_{Tidx} = \left( r_{i} - \textit{self}_{Sidx} \right) \bmod \textit{Node}_{cnt}
\end{equation}


Algorithm \ref{alg:per-round-map} defines a per-round mapping of nodes such that both sources and targets are paired within the same round. In this particular implementation a node will be paired with itself once every $n-rounds$. This can be leveraged as a checkpoint, reset, or indication to perform some other internal function.

\begin{algorithm}
\caption{CHRIP - Peer to Peer Mapping based on Round}
\label{alg:per-round-map}
\begin{algorithmic}[1]
\FOR{$\text{round}_{i}$ such that $0\leq i < n$}
\FOR {$\text{source}_{Sidx}$ such that $0 \leq Sidx < n$}
\STATE {$\text{source}_{Tidx} \leftarrow (\text{round}_{i} -  \textit{Source}_{idx}) \bmod \textit{Node}_{cnt} $}
\ENDFOR
\ENDFOR
\end{algorithmic}
\end{algorithm}

\begin{figure}[!h]
\centering
\includegraphics[width=3in]{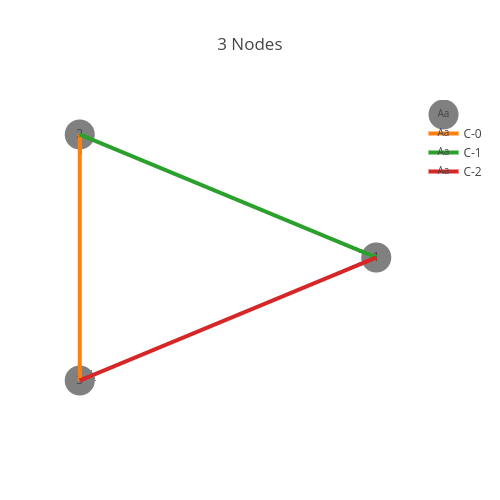}
\caption{Three rounds required to establish the three edges for a fully connected three-node graph.}
\label{fig:3-nodes}
\end{figure}

\begin{figure}[!h]
\centering
\includegraphics[width=3in]{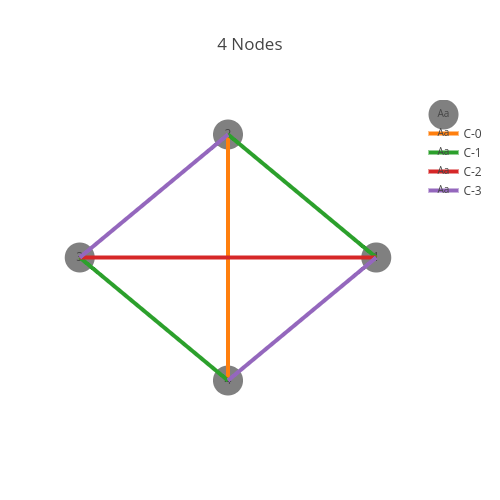}
\caption{The four rounds needed to create the 6 edges [$e= \frac{3*4}{2}$] for a fully connected four-node graph.}
\label{fig:4-nodes}
\end{figure}

\begin{figure}[!h]
\centering
\includegraphics[width=3in]{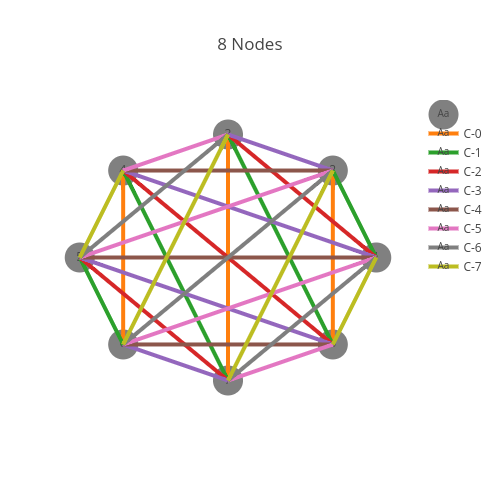}
\caption{The eight rounds needed to create the 28 edges [$e= \frac{7*8}{2}$] for a fully connected eight-node graph.}
\label{fig:8-nodes}
\end{figure}

Consider as three graphs show in Figures \ref{fig:3-nodes}-\ref{fig:8-nodes}. The three node graph is a trivial, yet critical example, and the four-node network shows how our relaxation of the minimum number of required rounds enables an extremely straight-forward communication protocol using the CHRIP algorithm. In each of the figures, edges of the graph are colored based on the round in which they were added.

While the trivial case, a graph with two nodes ($n=2$) is omitted for brevity. It should be obvious to the reader that a single round $r_0$ is needed connect two nodes. Shown in Fig. \ref{fig:3-nodes} are the three rounds needed to create a fully connected graph using the CHIRP pairing algorithm shown earlier. Figures \ref{fig:4-nodes} and \ref{fig:8-nodes} show the per round edge creation for four and eight nodes respectively.

\subsubsection{System Resilience to Node Loss} In order to measure the impact of a node loss on the system, let us consider the three items that any node requires to compute a target node index: node count, source index, and the current round. A loss of a node directly impacts the actual node count and removes a target node index from the global network. The change, would in any other system, also impact the future total number of rounds (i.e. one less node equates to one less round of communication), and it would also require a re-indexing of all the nodes with indexes greater than the node lost. 

\begin{figure}[!h]
\centering
\includegraphics[width=3in]{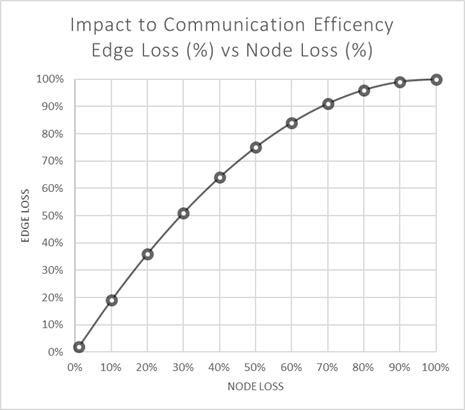}
\caption{Loss in communication efficiency (CE \%) as a function of Node Loss (\%) }
\label{fig:nodeloss}
\end{figure}

Rather than complicate the logic of the ultra-portable IoT devices, this implementation of CHIRP chooses to deal with the potentially short-term loss of communication efficiency by allowing the per node stored value of $\text{Node}_{cnt}$ to be greater than or equal the actual number of nodes in the system $\text{Node}_{actual}$. The impact on communication efficiency ($CE$) is defined as the ratio of missing nodes ($\text{Node}_{cnt}-\text{Node}_{actual}$ ) to per node stored node count ($\text{Node}_{cnt}$) as defined in Equation \ref{eq:eff}. A plot of the efficiency loss as a function of node loss is seen in Fig \ref{fig:nodeloss}. 

\begin{equation}\label{eq:eff}
CE = \frac{\text{Node}_{cnt}-\text{Node}_{actual}}{\text{Node}_{cnt}}
\end{equation}

\subsubsection{Resilience to Node (Re)Entry} Node (re)entry, from here on simply entry, into a system must satisfy two conditions:

\begin{enumerate}
\item The number of actual nodes within the system, prior to the addition of a new node, must be strictly less than the per-node expected maximum capacity:\\ $\text{Node}_{actual} < \text{Node}_{cnt}$
\item The index of the new node must not duplicate a node index that is already in the system
\end{enumerate}

If these two conditions are satisfied the final major hurdle for the entry of a node into an existing system is round index synchronization. Recall that in order to compute the per-round communication pair ($\text{source}_{Tidx}$), a node requires knowledge of the system's node capacity ($\text{Node}_{cnt}$), the current round ($\text{round}_{i}$), and of course its own in index ($\text{source}_{Sidx}$). Without knowledge of the current round the computation of the target index is not possible. Luckily, as seen in Equation \ref{eq:curr_round}, the current round is easily computed based on the node's own index ($\text{source}_{Sidx} $) as well as the index of the node currently attempting communication ($\text{source}_{Tidx}$).

\begin{equation}\label{eq:curr_round}
r_{i} = \left( \text{source}_{Tidx} + \text{source}_{Sidx} \right) \bmod \text{Node}_{cnt}
\end{equation}

A node entering an existing communication cycle could validate the current round index by tracking the node indices of several prior communication attempts and insuring the the equivalent round indices followed the expected sequential pattern. Table \ref{table:masked} shows the communication pattern of 7 of 8 known nodes during 8 rounds of communication. The actual round number is obscured as it would be to a node entering mid-cycle. Table \ref{table:extracted} shows the result of applying equation \ref{eq:curr_round} to the in-bound node index in order to determine the current round index. It should be noted that with some modifications, a similar approach could be used to adjust a nodes index (e.g. set a node index, if those communicating with you do no form a sequential set of round indices, modify your node index until they do).

\begin{table}[h!]
\centering
\caption{The view of 8 nodes per round communication targets, with node four and the current round index intentionally masked.}
\label{table:masked}
\begin{tabular}{@{}cllllllll@{}}
\toprule
                     & \multicolumn{8}{c}{Nodes}                                                                                                                                                                                                                      \\ \midrule
\textit{\textbf{$r_i$}} & \textit{\textbf{0}}       & \textit{\textbf{1}}       & \textit{\textbf{2}}       & \textit{\textbf{3}}                              & \textit{\textbf{4}} & \textit{\textbf{5}}       & \textit{\textbf{6}}       & \textit{\textbf{7}}       \\

\textit{\textbf{j}}  & \cellcolor[HTML]{C0C0C0}4 & 3                         & -                         & 1                                                & ?                   & 7                         & -                         & 5                         \\
\textit{\textbf{j+1}}  & 5                         & \cellcolor[HTML]{C0C0C0}4 & 3                         & 2                                                & ?                   & 0                         & 7                         & 6                         \\
\textit{\textbf{j+2}}  & 6                         & 5                         & \cellcolor[HTML]{C0C0C0}4 & -                                                & ?                   & 1                         & 0                         & -                         \\
\textit{\textbf{j+3}}  & 7                         & 6                         & 5                         & \cellcolor[HTML]{C0C0C0}{\color[HTML]{333333} 4} & ?                   & 2                         & 1                         & 0                         \\ 
\textit{\textbf{j+4}}  & -                         & 7                         & 6                         & 5                                                & ?                   & 3                         & 2                         & 1                         \\
\textit{\textbf{j+5}}  & 1                         & 0                         & 7                         & 6                                                & ?                   & \cellcolor[HTML]{C0C0C0}4 & 3                         & 2                         \\
\textit{\textbf{j+6}}  & 2                         & -                         & 0                         & 7                                                & ?                   & -                         & \cellcolor[HTML]{C0C0C0}4 & 3                         \\
\textit{\textbf{j+7}}  & 3                         & 2                         & 1                         & 0                                                & ?                   & 6                         & 5                         & \cellcolor[HTML]{C0C0C0}4 \\
\bottomrule
\end{tabular}
\end{table}

\begin{table}[h]
\centering
\caption{The inbound node indices in conjunction with the known Node Count is sufficent to compute the current round index and validate that they are progressing in the correct sequence. }
\label{table:extracted}
\begin{tabular}{@{}ccl@{}}
\textit{\textbf{Unknown $R_i$}} & \textit{\textbf{In-bound Node Idx}} & \textit{\textbf{Computed Round Idx}} \\ \toprule
\textit{\textbf{j}}          & 0                                   & (4 + 0) mod 8 = 4                    \\ \midrule
\textit{\textbf{j+1}}        & 1                                   & (4 + 1) mod 8 = 5                    \\
\textit{\textbf{j+2}}        & 2                                   & (4 + 2) mod 8 = 6                    \\
\textit{\textbf{j+3}}        & 3                                   & (4 + 3) mod 8 = 7                    \\
\textit{\textbf{j+4}}        & None (self-loop)                       & (4 + 4) mod 8 = 0                    \\
\textit{\textbf{j+5}}        & 5                                   & (4 + 5) mod 8 = 1                    \\
\textit{\textbf{j+6}}        & 6                                   & (4 + 6) mod 8 = 2                    \\
\textit{\textbf{j+7}}        & 7                                   & (4 + 7) mod 8 = 3                    \\ \bottomrule
\end{tabular}
\end{table}

\subsection{Enabling Secure Communication S-CHIRP}

The simplest way to enhance communication within CHIRP at scale is to to introduce a permutation in the rounds ordering. While this mechanism is insufficient at low node counts there are solutions that could be utilized in small ($\leq8$ node) networks - specifically per-node keys. Table \ref{table:node_permutations} shows the rapid increase in potential round permutations given increasing node counts. 

\begin{table}[h!]
\centering
\caption{The number of possible round permutations for a given number of nodes (and thus rounds).}
\label{table:node_permutations}
\begin{tabular}{@{}cc@{}}
\toprule
Nodes / Rounds & Possible Round Permutations         \\ \midrule
8          & $4.03 \times 10^5$   \\ 
10         & $3.63 \times 10^6$   \\
16         & $2.09 \times 10^{13}$  \\
32         & $2.63 \times 10^{35}$  \\
64         & $1.26 \times 10^{89}$  \\
128        & $3.85 \times 10^{215}$ \\ \bottomrule
\end{tabular}
\end{table}

Consider a standard cycle ($C$) in CHIRP algorithm as a monotonically increasing set of round indices from zero to the maximum number of nodes minus 1 as show in equation \ref{eq:cycle}. A unique permutation of these $\text{Node}_{cnt}$ elements within cycle $C$ results in a shuffling of communication between the nodes in a system that requires immense resources to track, detect and exploit. 

\begin{equation}\label{eq:cycle}
C = \left[0,1,2, \cdots, \text{Node}_{cnt}-2, \text{Node}_{cnt}-1\right]
\end{equation}

While different approaches exist to synchronize permutations across a set of nodes this work suggests two different approaches, each with their own challenges and benefits. 

The first option is using the classic Fisher-Yates Shuffle \cite{fisher1963statistical} \cite{durstenfeld1964algorithm} (or Knuth shuffle \cite{knuth1997art}) algorithm\footnote{It is also interesting to note that Sattolo's algorithm \cite{sattolo1986algorithm} could also be substituted for the Fisher-Yates shuffle since it directly creates random cyclic permutations.}. The fundamental advantages to the Fisher-Yates shuffle include: $\mathcal{O}(n)$ time complexity and the ability to perform the permutation in-place without any additional storage requirements. The massive caveat to this approach is the need for a random number generator that is consistent across a heterogeneous swarm of devices. Should this caveat be addressed, a permuted cycle ($Pr(C)$) could be easily computed within each device based on some seed value. Changing the network that a node communicates on is as simple as resetting the seed value for the permutation.

The second option is the off-device creation of a permutation through any desired method ranging from "hand-crafted" permutations to "randomized" permutations. In terms of benefit, the process of iterating through a stored array of indices is fast, reusable, and accessible across heterogeneous devices. The disadvantage is the upfront storage costs for a $\text{Node}_{cnt}$ length array, especially as the number of nodes increases. This is likely a non-issue in all but extreme cases, as even a 10,000 node system would only require 40 KB to store an entire cycle permutation (a million nodes? 4 MB).  

Irrespective of the mechanism used to deliver, store, and iterate through the round permutations the impact to the four requirements presented in the prior section is highlighted below. The assumption is that the cycle permutation $Pr(C)$ is index addressable between $\left[0,\text{Node}_{cnt}\right)$.

\subsubsection{Per-Round Bi-directional Secure Communication}
The minimum number of rounds needed to accomplish complete  \textbf{secure} pair-wise communication is still $R_{min} = n$. S-CHIRP completes  \textbf{secure} pairwise communication within $n$ rounds, where each round contains at least $\frac{n-1}{2}$ distinct pairs of communicating nodes.  In order to determine the index of a source node's "target node pair" ($target_{idx}$) one additional piece of information is required: an array containing the permutation of the cycle rounds ($Pr(C)$). 

With four parameters - any node can be securely paired with any other node using simple modular arithmetic during any round using equation \ref{eq:spairs}. 

\begin{equation}\label{eq:spairs}
\textit{self}_{Tidx} = \left( Pr(C)[r_{i}] - \textit{self}_{Sidx} \right) \bmod \textit{Node}_{cnt}
\end{equation}

Algorithm \ref{alg:per-round-smap} defines a per-round mapping of nodes such that both sources and targets are paired within the same round, but there is no distinct linear sweep through the offset between pairs. See Table \ref{table:nosweep} for an example of the per-round communication for a permuted 8 round cycle $Pr(C)=[7,5,2,0,4,6,1,3]$, and then see Table \ref{table:masked} diagonals for an example of this sweep in the insecure version with a simple 5 round offset, $C = [5,6,7,0,1,2,3]$. 

\begin{algorithm}
\caption{CHRIP - Peer to Peer Mapping based on Round}
\label{alg:per-round-smap}
\begin{algorithmic}[1]
\STATE {$Pr(C) \Leftarrow Permutation([0,1,\cdots,\textit{Node}_{cnt}-1] $}
\FOR{$\text{r}_{i}$ such that $0\leq i < n$}
\FOR {$\text{source}_{Sidx}$ such that $0 \leq Sidx < n$}
\STATE {$\text{source}_{Tidx} \leftarrow (Pr(C)[\text{r}_{i}] -  \textit{Source}_{idx}) \bmod \textit{Node}_{cnt} $}
\ENDFOR
\ENDFOR
\end{algorithmic}
\end{algorithm}

\begin{table}[h!]
\centering
\caption{A permutation of an 8-round cycle yields no discernible "sweeps" through node offsets - generally recognized as diagonals within the matrix.}
\label{table:nosweep}
\begin{tabular}{@{}ccclllllll@{}}
\toprule
   & \textit{\textbf{}} & \multicolumn{8}{c}{\textit{\textbf{Nodes}}}       \\ \midrule
Ri & Pr(C)              & 0                     & 1 & 2 & 3 & 4 & 5 & 6 & 7 \\
0  & 7                  & 6                     & 5 & 4 & - & 2 & 1 & 0 & - \\
1  & 5                  & 4                     & 3 & - & 1 & 0 & 7 & - & 5 \\
2  & 2                  & 1                     & 0 & 7 & 6 & 5 & 4 & 3 & 2 \\
3  & 0                  & 7                     & 6 & 5 & 4 & 3 & 2 & 1 & 0 \\
4  & 4                  & 3                     & 2 & 1 & 0 & 7 & 6 & 5 & 4 \\
5  & 6                  & \multicolumn{1}{l}{5} & 4 & 3 & 2 & 1 & 0 & 7 & 6 \\
6  & 1                  & \multicolumn{1}{l}{-} & 7 & 6 & 5 & - & 3 & 2 & 1 \\
7  & 3                  & \multicolumn{1}{l}{2} & - & 0 & 7 & 6 & - & 4 & 3 \\ \bottomrule
\end{tabular}
\end{table}

\section{Conclusion \& Future Work}\label{FUTURE}

This work presents a mechanism to enable secure peer to peer communication in ultra-portable, heterogeneous devices using an extremely simple protocol which can be enhanced at low cost to segregate networks and prevent un-sanctioned communication. Peer-to-Peer communication requires knowledge of a key (which can be transmitted using any one of a number of existing strategies), and knowledge of two pieces of information the current cycle, and an internal address. The minimum time to communicate to all nodes in a key-locked round-robin N-node network is N-rounds. Future work includes multi-agent simulation and targeted domain applications using a variety of communication mediums and existing technologies. Additionally,this simple protocol could be used as a wrapper to IPv6 enabling communication along entire specific subnet masks without requiring individual nodes to have knowledge of network topology.

\bibliographystyle{IEEEtran}
\bibliography{swarm.bib}
%

\end{document}